\documentclass[aps,preprint,onecolumn,showpacs]{revtex4}

\usepackage{graphicx}% Include figure files
\usepackage{dcolumn}% Align table columns on decimal point
\usepackage{bm}% bold math

\begin{document}

\title{The Anti-Cloak}

\author{Huanyang Chen$^*$, Xudong Luo, and
Hongru Ma}

\affiliation{Department of Physics, Shanghai Jiao Tong University,
Shanghai 200240, China}

\author{C.T. Chan}

\affiliation{Department of Physics, The Hong Kong University of
Science and Technology, Clear Water Bay, Hong Kong, China}

\begin{abstract}
A kind of transformation media, which we shall call the
``anti-cloak'', is proposed to partially defeat the cloaking effect
of the invisibility cloak. An object with an outer shell of
``anti-cloak'' is visible to the outside if it is coated with the
invisible cloak. Fourier-Bessel analysis confirms this finding by
showing that external electromagnetic wave can penetrate into the
interior of the invisibility cloak with the help of the anti-cloak.
\end{abstract} \maketitle

%%%%%%%%%%%%%%%%%%%%%%%%%%  body  %%%%%%%%%%%%%%%%%%%%%%%%%%

The ideas of transformation optics and cloaking [1-4] have attracted
keen interest both in theory [1-18] and in experiment [19]. The
cloaking effect has been proved using different methods, such as ray
tracing [8], full wave simulation employing finite element methods
[9] and the finite difference time domain methods [10], and the Mie
scattering models [11,12]. In particular, Ruan et al. [11] employed
Mie scattering models to confirm that a cylindrical cloak with the
ideal material parameters is indeed a perfect invisibility cloak
using Fourier-Bessel analysis. Similar approaches [12] were used to
confirm the perfect cloaking effect of the spherical cloak. However,
essentially all the aforementioned examples that demonstrated the
perfect cloaking effect did not consider embedded objects with
material anisotropy inside the cloak. In this paper, we show that
the perfect cloaking effect can be defeated by adding another kind
of transformation media inside the cloak (i.e., the anisotropy
inside the cloak is considered, see in Fig. 1 schematically). We
shall construct an example which demonstrates that an object with an
outer shell of a specific form of negative index anisotropic
material cannot be made entirely invisible by the transformation
media cloak.

\begin{figure}
\begin{center}
\includegraphics[angle=-0,width=1.0\columnwidth] {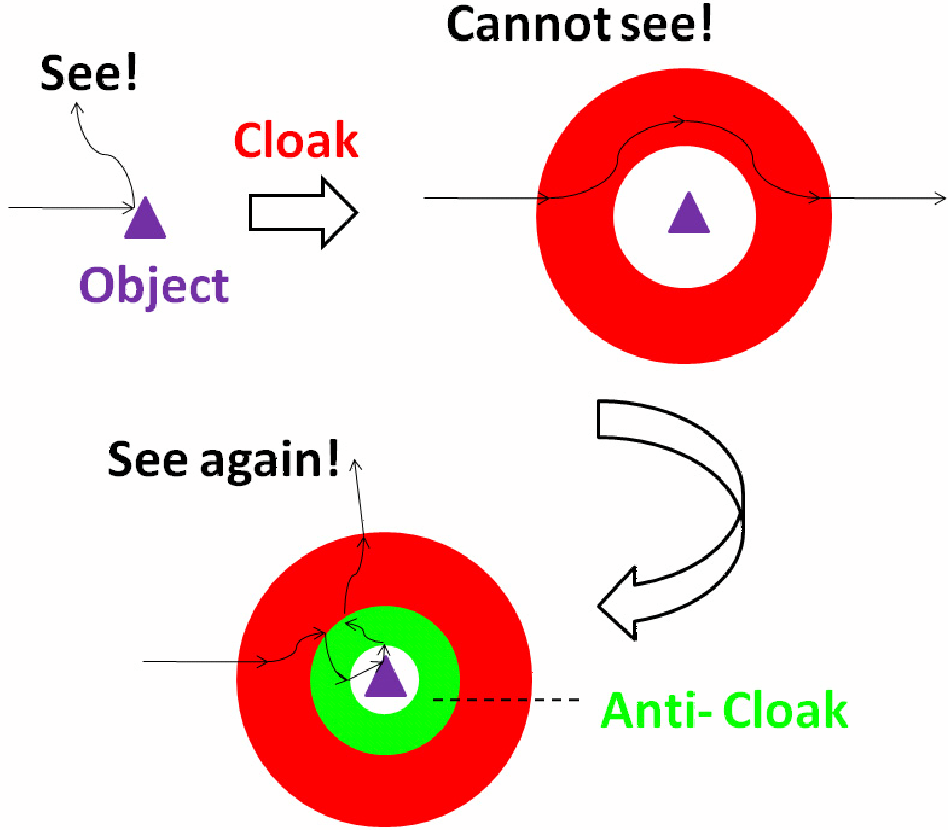}
\end{center}
\caption{(Color online) The schematic figure to illustrate the
cloaking effect and anti-cloaking effect.}\label{fig.1}
\end{figure}

\begin{figure}
\begin{center}
\includegraphics[angle=-0,width=1.0\columnwidth] {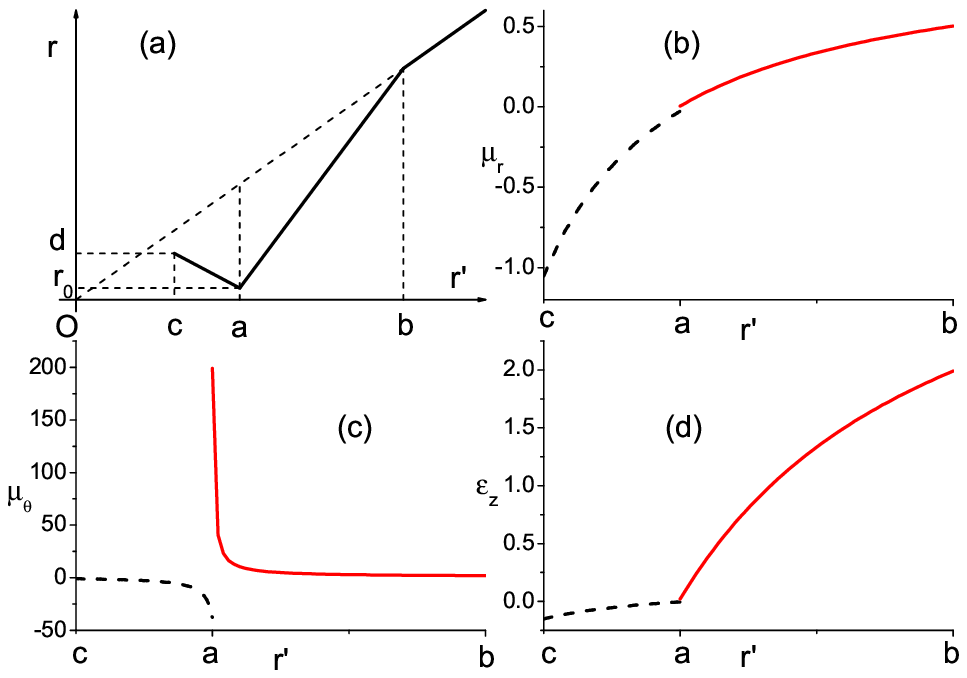}
\end{center}
\caption{(Color online) (a) The coordinate transformation of the
cloak ($a < r' < b)$ and anti-cloak ($c < r' < a)$. (b) $\mu _r $
for cloak and anti-cloak. (c) $\mu _\theta $ for cloak and
anti-cloak. (d) $\varepsilon _z $ for cloak and anti-cloak. The red
solid lines in b-d denote the parameters of cloak, while the black
dashed lines in b-d denote the parameters of
anti-cloak.}\label{fig.2}
\end{figure}

Starting from the mapping [6, 14] (see in Fig. 2(a)), $r = \frac{b -
r_0 }{b - a}(r' - b) + b,$ the required parameters for a partial
cylindrical cloak (transverse electric (TE) mode is considered here)
are obtained as follow,

\begin{equation}
\label{eq1} \mu _r = \frac{r' - a_1 }{r'},\;\mu _\theta =
\frac{r'}{r' - a_1 },\;\varepsilon _z = (\frac{b - r_0 }{b -
a})^2\frac{r' - a_1 }{r'},
\end{equation}

\noindent where $a_1 = \frac{a - r_0 }{b - r_0 }b.$ This partial
cloak can reduce the total scattering cross section of a perfect
electrical conductor (PEC) cylinder from its radius $r' = a$ to an
equivalent PEC cylinder whose radius is $r = r_0 $. In the limit as
$r_0 $ goes to zero, the partial cloak becomes perfect [1].

Now let us add another coordinate transformation inside the cloak
($c < r' < a)$ as depicted in Fig. 2(a), $r = \frac{d - r_0 }{c -
a}(r' - c) + d.$ The corresponding material parameters are then,

\begin{equation}
\label{eq2} \mu _r = \frac{r' - a_2 }{r'},\;\mu _\theta =
\frac{r'}{r' - a_2 },\;\varepsilon _z = (\frac{d - r_0 }{c -
a})^2\frac{r' - a_2 }{r'},
\end{equation}

\noindent where $a_2 = \frac{ad - cr_0 }{d - r_0 }.$ We note that
these values are negative. We call this kind of transformation media
the "anti-cloak" as we shall see that they cancel partially the
effect of an invisibility cloak.

In the same spirit of the partial cloak, when a PEC cylinder with a
radius $r' = c$ is coated with the anti-cloak in direct contact with
the partial cloak, the total scattering cross section will be
changed into that of an equivalent PEC cylinder whose radius is $r =
d$. We note that there are no PEC boundary between the cloak and the
anti-cloak (at $r' = a$), they are in direct contact.

Doing the same Fourier-Bessel analysis in [11], we can obtain the
electric fields in each region,

\begin{equation}
\label{eq3}
\begin{array}{l}
 (b \le r):E_z = \sum\limits_l {\alpha _l^{in} J_l (k_0 r)\exp (il\theta )}
\\
 \mbox{ } + \alpha _l^{sc} H_l (k_0 r)\exp (il\theta ), \\
 (a \le r < b):E_z = \sum\limits_l {\alpha _l^1 J_l (k_1 (r - a_1 ))\exp
(il\theta )} \\
 \mbox{ } + \alpha _l^2 H_l (k_1 (r - a_1 ))\exp (il\theta ), \\
 (c \le r < a):E_z = \sum\limits_l {\alpha _l^3 J_l (k_2 (r - a_2 ))\exp
(il\theta )} \\
 \mbox{ } + \alpha _l^4 H_l (k_2 (r - a_2 ))\exp (il\theta ). \\
 \end{array}
\end{equation}

\noindent where $J_l \backslash H_l $ are the $l$-order
Bessel$\backslash $Hankel function of the 1st kind, $k_0 $ is the
wave vector of the light in vacuum, $k_1 = \frac{b - r_0 }{b - a}k_0
,\;k_2 = \frac{d - r_0 }{c - a}k_0 ,$ $\alpha _l^{in} $ and $\alpha
_l^{sc} $ are the incident and scattering coefficients outside the
cloak, $\alpha _l^i (i = 1,2,3,4)$ are the expansion coefficients
for the field in the cloak and anti-cloak. The primes are dropped
for aesthetic reasons from here. From the continuous boundary
conditions (at $r = b$ and $r = a$) and the PEC boundary ($E_z = 0$
at $r = c$), we can obtain that,

\begin{equation}
\label{eq4}
\begin{array}{l}
 \alpha _l^1 = \alpha _l^3 = \alpha _l^{in} , \\
 \alpha _l^2 = \alpha _l^4 = \alpha _l^{sc} , \\
 \alpha _l^{sc} = - \frac{J_l (k_0 d)}{H_l (k_0 d)}\alpha _l^{in} . \\
 \end{array}
\end{equation}

\noindent This result confirms that the PEC cylinder with its radius
$r = c$ coated with the anti-cloak and cloak is equivalent to a PEC
cylinder with its radius $r = d$ in the view of outside world.

\begin{figure}
\begin{center}
\includegraphics[angle=-0,width=0.45\columnwidth] {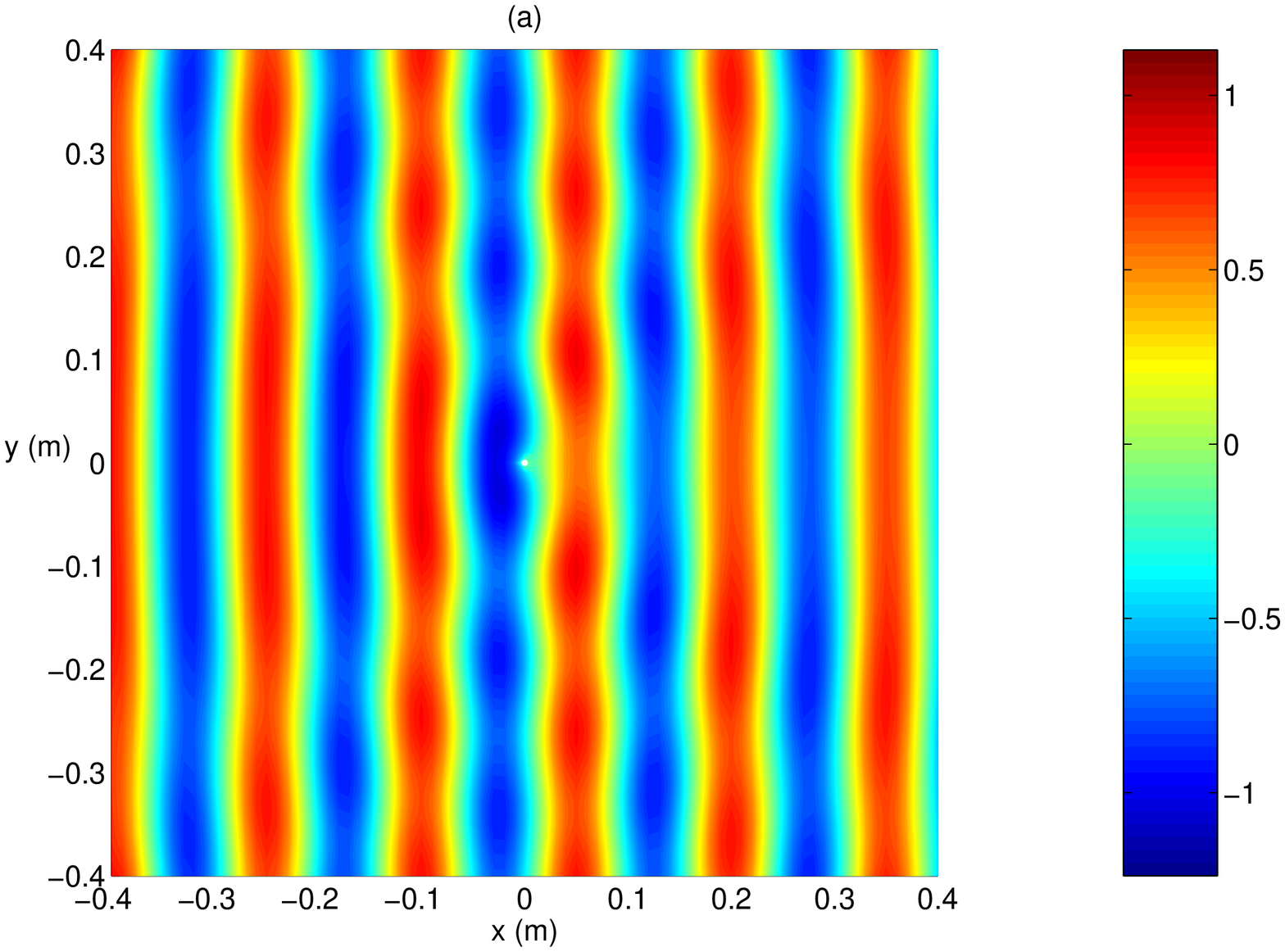}
\includegraphics[angle=-0,width=0.45\columnwidth] {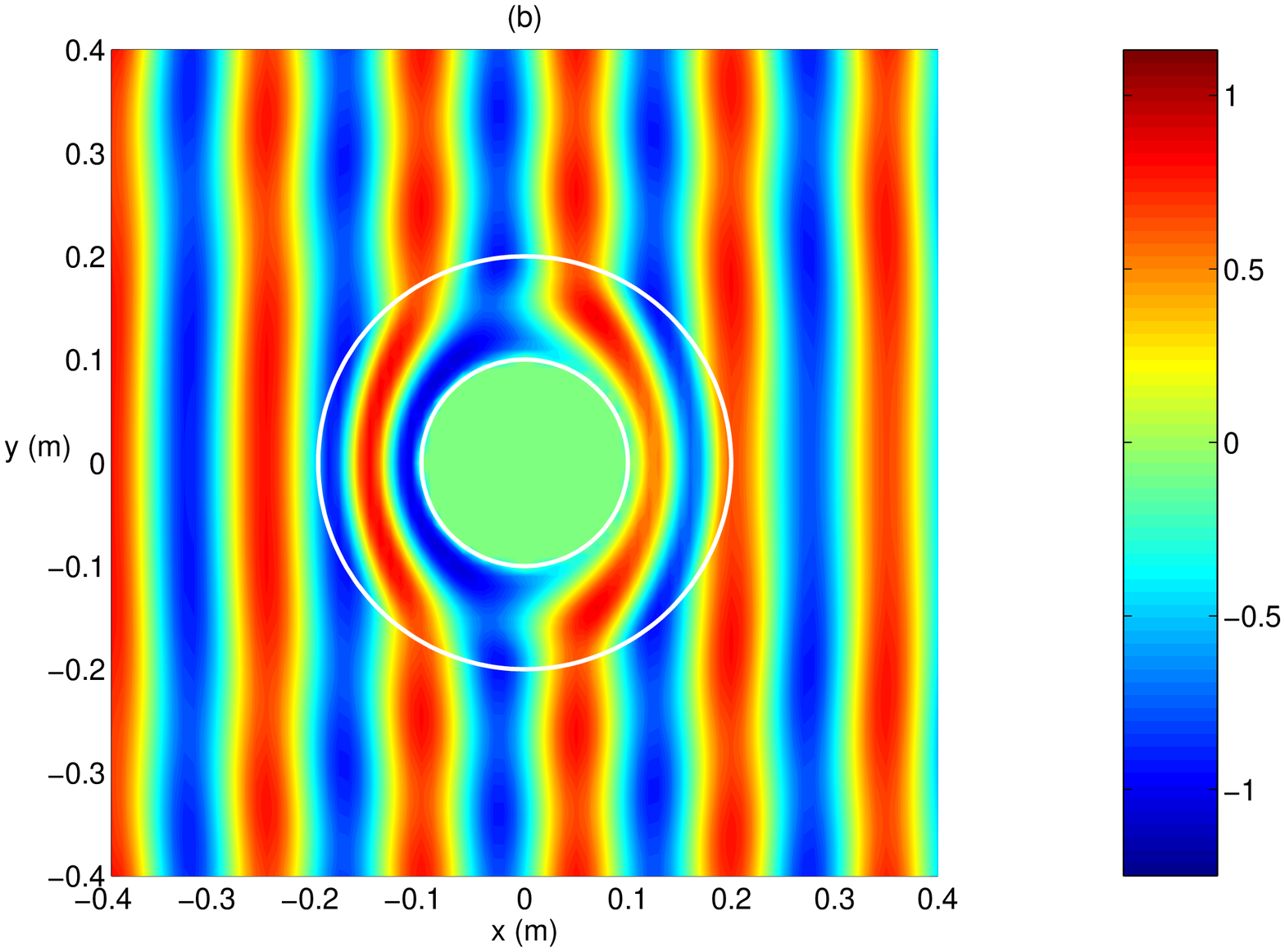}\\
\includegraphics[angle=-0,width=0.45\columnwidth] {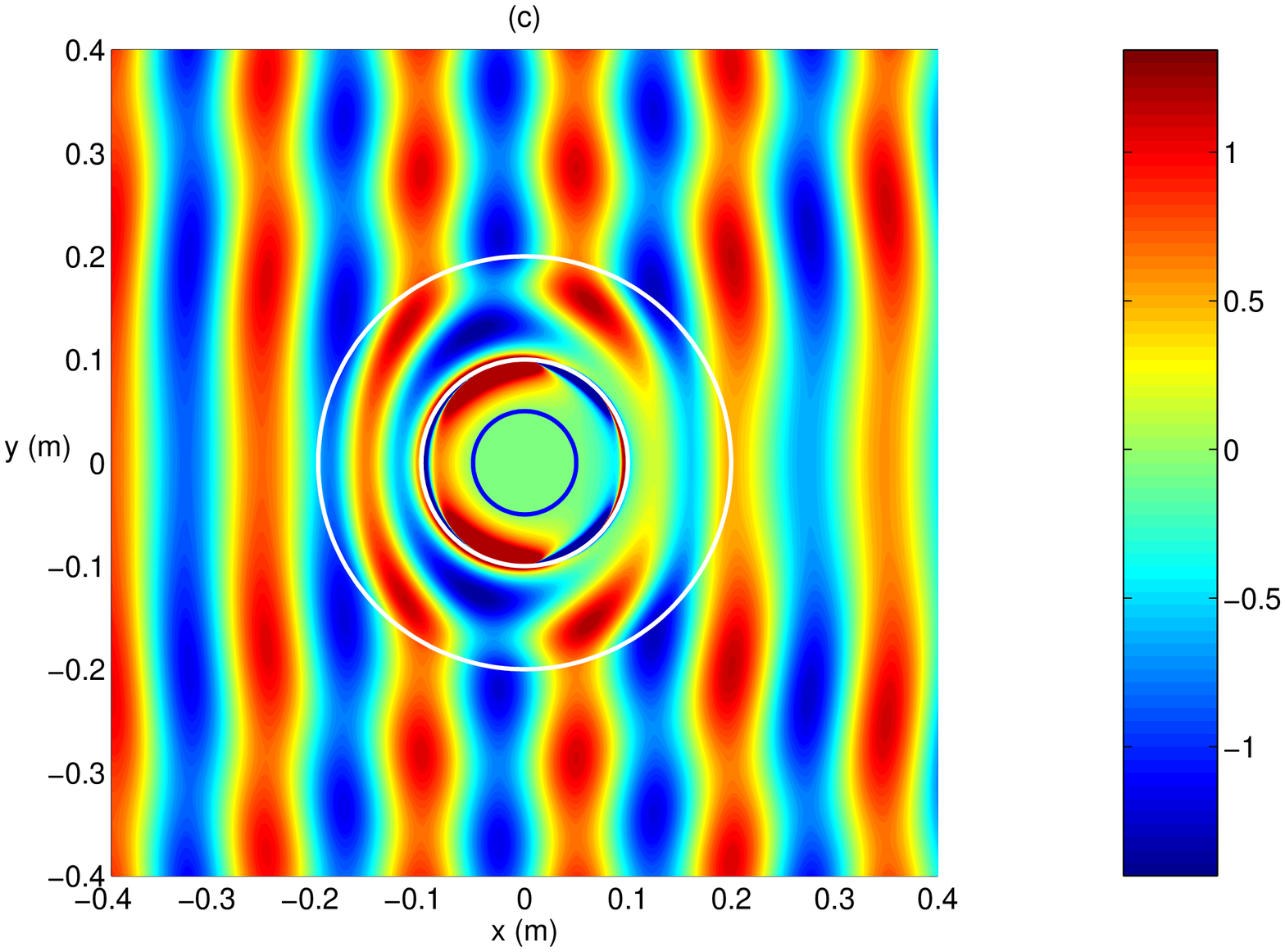}
\includegraphics[angle=-0,width=0.45\columnwidth] {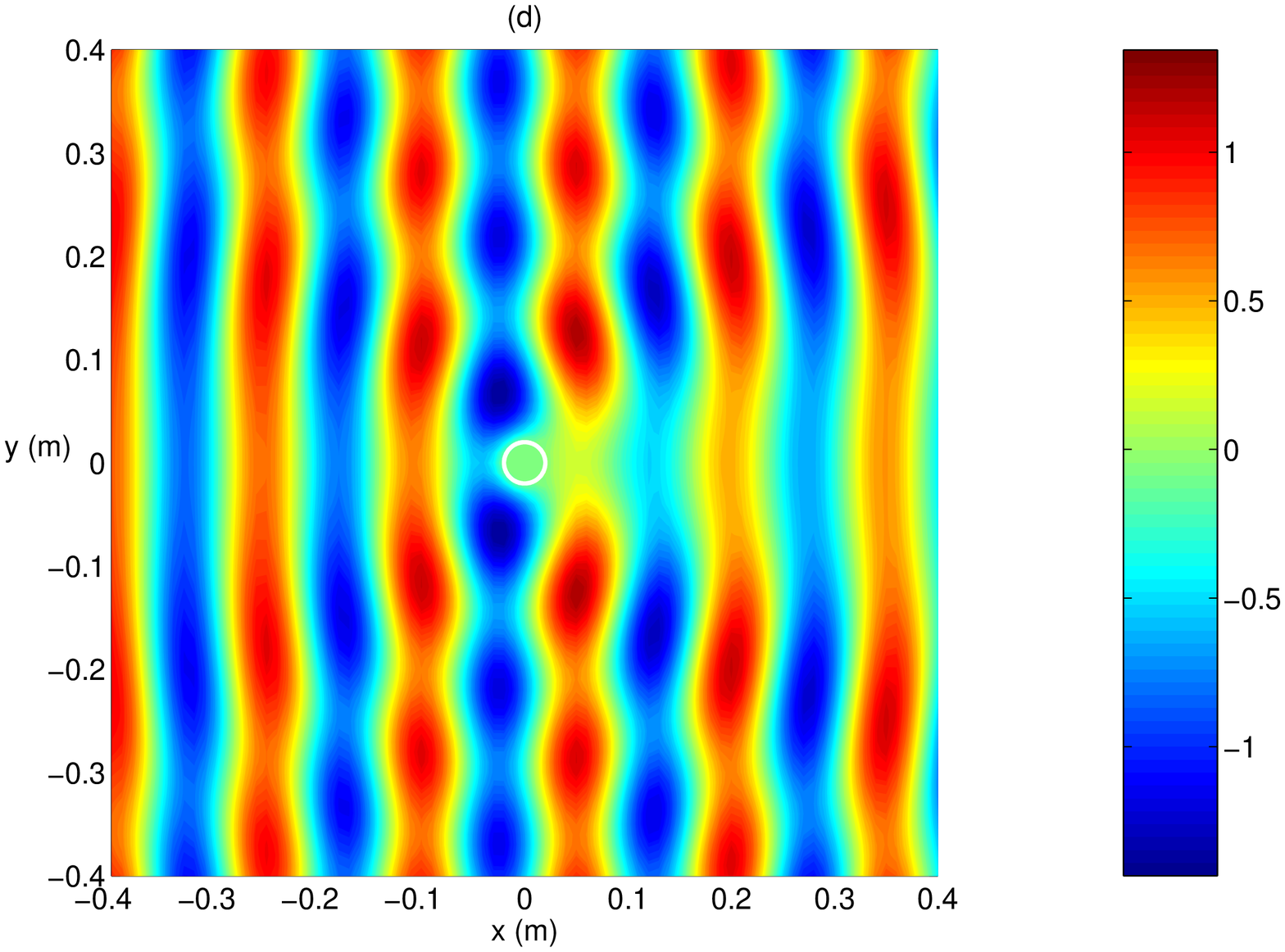}
\end{center}
\caption{(Color online) The electric filed distribution for, (a) a
tiny PEC cylinder with a radius $r_0 $(outlined by the write point);
(b) a PEC cylinder with a radius $a$ wearing a partial cloak (the
outer radius is $b$, the inner and outer boundaries of the cloak are
outlined by the write solid lines); (c) a PEC cylinder with a radius
$c$ wearing an anti-cloak and the same partial cloak as in (b) (the
inner and outer boundaries of the cloak are outlined by the write
solid lines while the inner boundary of the anti-cloak at $r = c$ is
outlined by the blue solid line); (d) a PEC cylinder with a radius
$d$(outlined by the write solid line).}\label{fig.3}
\end{figure}

To demonstrate the properties of the anti-cloak, we set $a = 0.1m$,
$b = 0.2m$, $c = 0.05m$, $d = 0.02m$, $r_0 = 0.001m$. We plot the
parameters of the cloak and anti-cloak at different radial positions
in Fig. 2(b)-(d). All the parameters of anti-cloak are negative
because of the negative slope of the coordinate transformation. A
plane wave is incident from left to right with the frequency $2GHz$.
In Fig. 3(a), we plot the scattering pattern of a PEC cylinder with
a radius $r_0 $. The tiny PEC cylinder causes little scattering for
the incoming plane wave which can be treated as almost invisible. In
Fig. 3(b), we plot the scattering pattern of a PEC cylinder with a
radius $a$ coated by a partial cloak. The outer radius of the cloak
is $b$. We see that the partial cloak reduces substantially the
scattering of the PEC cylinder with its radius $a$ when we compare
Fig. 3(a) and Fig. 3(b). When $r_0 $ is made as small as we like,
the scattering becomes vanishing small. In Fig. 3(c), we plot the
scattering pattern of a PEC cylinder with a radius $c$ coated by an
anti-cloak and a partial cloak [20]. The anti-cloak is located in $c
< r < a$, the cloak locates in $a < r < b$. Without the anti-cloak,
the wave basically goes around the shielded region, but if the
anti-cloak is in contact with the cloak, EM wave from outside can go
into the anti-cloak to interact with the object inside. The
scattering of the cloak is enlarged again to that of an equivalent
PEC cylinder whose radius is $d$. We plot the scattering pattern of
the equivalent PEC cylinder in Fig. 3(d). When $c = d$, the
anti-cloak together with the partial cloak becomes invisible, that
means one can directly see the PEC cylinders with radius $r = c$,
and the anti-cloak cancels out the effect of the partial cloak
completely. For aesthetic reasons, if the electric field is larger
than the maximum value in color bar in Fig. 3(c), we have replaced
this overvalued field with the maximum value when plotting Fig.
3(c). If the electric field is smaller than the minimum value, we
have replaced this overvalued field with the minimum value when
plotting Fig. 3(c).

\begin{figure}
\begin{center}
\includegraphics[angle=-0,width=0.6\columnwidth] {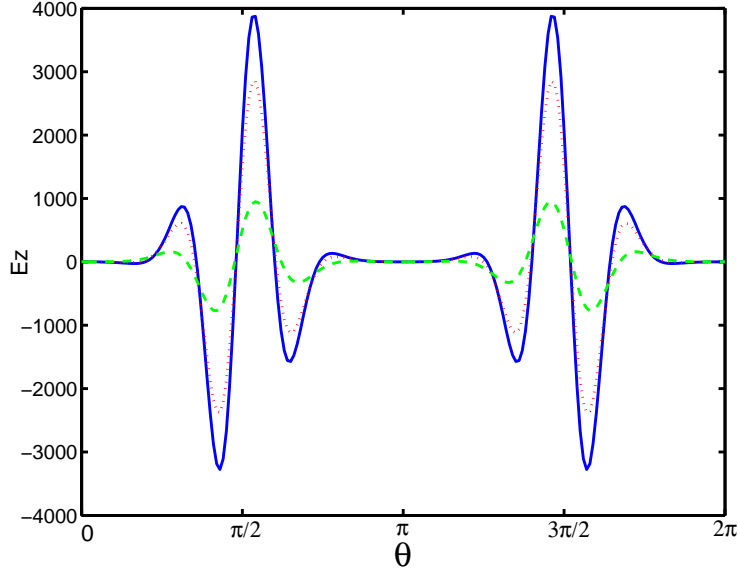}
\end{center}
\caption{(Color online) The electric field distribution at $r = a -
0.1r_0 $ (red dashed line), $r = a$ (solid blue line) and $r = a +
0.1r_0 $ (green dotted line).}\label{fig.4}
\end{figure}

Due to the continuous coordinate transformation at $r = a$, the
impedances are matched at this touching boundary of the cloak and
anti-cloak. The electric field is very large at this touching
boundary. To show this property, we plot the electric field for
different angles at fix radii near $r = a$ in Fig. 4. Three fixed
radii are chosen, $r = a - 0.1r_0 $, $r = a$ and $r = a + 0.1r_0 $.
We find that the electric field near $r = a$ is very large.

Analytically, one can obtain the electric field at $r = a$ as
follow,

\begin{equation}
\label{eq5}
\begin{array}{l}
 E_z = \sum\limits_l {\alpha _l^{in} J_l (k_0 r_0 )\exp (il\theta ) + \alpha
_l^{sc} H_l (k_0 r_0 )\exp (il\theta )} \\
 = \sum\limits_l {\alpha _l^{in} \exp (il\theta )[J_l (k_0 r_0 ) - \frac{J_l
(k_0 d)}{H_l (k_0 d)}H_l (k_0 r_0 )]} . \\
 \end{array}
\end{equation}

\noindent The term $H_l (k_0 r_0 )$ becomes very large when $r_0 $
is small, that is why we obtain large electric field above.

Since we can make $r_0 $ as small as we like, we reach the
conclusion that an almost perfect cloak can be defeated by an
anti-cloak. In other words, the transformation media cloak is not a
panacea as there exists some objects that it cannot hide. In the
limit that $r_0 $ is exactly zero, the situation requires further
mathematical analysis due to the singularity properties of the
anti-cloak and cloak ($H_l (k_0 r_0 )$ diverges when $r_0 $ goes to
zero). From a physical standpoint, we may argue as follows. Near the
inner boundary of the invisibility cloak, $\mu _r $ goes to zero and
$\mu _\theta $ goes to infinity and they are positive, while near
the outer boundary of the anti cloak, $\mu _r $ goes to zero and
$\mu _\theta $ goes to infinity from the negative side. The positive
singular values have to come from an in-phase resonance while the
negative infinity comes from out-of-phase resonance. If we put them
in contact, the system response is canceled out, and the cloaking
effect is weaken or even destroyed. The surface mode resonance at $r
= a$ is excited and contributes to the large electric field. In
addition, if the losses are considered, the electric field will
become finite for $r_0 $ is exactly zero. The cylindrical anti-cloak
concept could be extended to three dimensions.

In conclusion, we find that the invisible cloak cannot hide the
enclosed domain if the inside domain has a shell of anti-cloak. The
properties are demonstrated by using the Fourier-Bessel analysis and
finite-element full wave simulations. The anti-cloak region is an
anisotropic negative refractive shell that is impedance matched to
the cloak outside, which has a positive refractive index. It is
known that [21] a negative refractive index medium ``cancels'' the
space of a positive index medium that has the same impedance. So, a
heuristic way of understanding the operation of an anti-cloak is
that it annihilates the functionality of the interior part of the
invisibility cloak, and effectively shifts the enclosed PEC region
outwards to make contact with the outer part of the cloaking shell
that is not ``canceled''. This leads to a finite cross section.

\section*{Acknowledgments}
This work was supported by the National Natural Science Foundation
of China under grand No.10334020 and in part by the National
Minister of Education Program for Changjiang Scholars and Innovative
Research Team in University, and Hong Kong Central Allocation Fund
HKUST3/06C.\\
$^*$Correspondence should be addressed to: kenyon@ust.hk
%%%%%%%%%%%%%%%%%%%%%%% References %%%%%%%%%%%%%%%%%%%%%%%%%


\begin{thebibliography}{99}
\bibitem{20foot} We have run simulations with a finite-element code, and obtained the
same results as in Fig. 3(c).
\end{thebibliography}
\end{document}